\documentclass[nofootinbib,preprintnumbers,showpacs,noshowkeyws]{revtex4}

\usepackage{epsfig}

\begin{document}
\title{Higgs quartic coupling and neutrino sector evolution in 2UED models}

\author{Ammar Abdalgabar}
\email[Email: ]{ammar.abdalgabar@students.wits.ac.za}
\affiliation{National Institute for Theoretical Physics; School of Physics, University of the Witwatersrand, Wits 2050, South Africa}
\author{A.~S.~Cornell}
\email[Email: ]{alan.cornell@wits.ac.za}
\affiliation{National Institute for Theoretical Physics; School of Physics, University of the Witwatersrand, Wits 2050, South Africa}
\author{Aldo Deandrea}
\email[Email: ]{deandrea@ipnl.in2p3.fr}
\affiliation{Universit\'e de Lyon, F-69622 Lyon, France; Universit\'e Lyon 1, CNRS/IN2P3, UMR5822 IPNL, F-69622 Villeurbanne Cedex, France}
\author{Ahmad Tarhini}
\email[Email: ]{tarhini.ahmad@gmail.com}
\affiliation{King Abdullah University of Science and Technology (KAUST), Division of Physical Science and Engineering, Thuwal 23955-6900, Saudi Arabia.}

\begin{abstract}
Two compact universal extra dimensional models are an interesting class of models for different theoretical and phenomenological 
issues, such as the justification of having three standard model fermion families, suppression of proton decay rate, dark matter parity from relics of the 6-dimensional Lorentz symmetry, origin of masses and mixings in the standard model. However, these theories are merely effective ones, with typically a reduced range of validity in their energy scale. We explore two limiting cases of the three standard model generations all propagating in the bulk or all localised to a brane, from the point of view of renormalisation group equation evolutions for the Higgs sector and for the neutrino sector of these models. The recent experimental results of the Higgs boson from the LHC allow, in some scenarios, stronger constraints on the cut-off scale to be placed, from the requirement of the stability of the Higgs potential.
\end{abstract}

\date{27 July, 2013}
\preprint{LYCEN 2013-07, WITS-CTP-118}
\maketitle


\section{Introduction}
\label{sec:1}

\par The Higgs mechanism has intrigued both theorists and experimentalists for a long time, being one of the central pieces in the construction of the Standard Model (SM) and its extensions. The Large Hadron Collider (LHC) was indeed built to discover the missing pieces of the SM and to search for new particles Beyond the Standard Model (BSM). The ATLAS and CMS experiments announced last year the discovery of a Higgs particle compatible with the SM expectations with a mass of about 126 GeV \cite{ATLAS:2012zz,CMS:2012zz}, and updated results with more data have been recently released (see for example \cite{ATLAS-CONF-2013-014,Chatrchyan:2013lba}). As we increase the energy scale above the Electro-Weak (EW) scale, the quartic couplings may become smaller and eventually become negative, where as a consequence, the potential becomes unbounded from below and the EW vacuum becomes unstable. In the context of the SM this problem was investigated a long time ago (see for example \cite{Sher:1988mj}) and continuously reinvestigated till recently (see for example \cite{Degrassi:2012ry}) as Higgs data and more precise calculations became available. 

\par This is also relevant for BSM and in particular those we shall consider in the following, as this can give bounds on the limit of validity of the effective theory. Among the models which explore new physics that may operate near the TeV scale, those with extra spatial dimensions  \cite{Antoniadis:1990ew} open many possibilities for model building which can be constrained \cite{Appelquist:2000nn} or explored in the near future. In this context there have been many attempts to understand the origin of masses and their mixings by using the Renormalisation Group Equations (RGEs) particularly for the Universal Extra Dimension (UED) models and their possible extensions (see for example \cite{Cornell:2012qf} and references therein). In this case the problem of vacuum stability and the triviality of the Higgs potential can be quite relevant as running effects are more pronounced in these models (with respect to the SM). For some recent work concerning 5 and 6-dimensional UED models see \cite{Kakuda:2013kba}. 

\par In particular, different models with two extra-dimensions have been proposed, such as $T^2/Z_2$ \cite{Appelquist:2000nn}, the chiral square $T^2/Z_4$ \cite{Dobrescu:2004zi, Burdman:2005sr}, $T^2/(Z_2 \times Z'_2)$ \cite{Mohapatra:2002ug}, $S^2/Z_2$ \cite{Maru:2009wu}, the flat real projective plane $RP^2$ \cite{Cacciapaglia:2009pa}, and the real projective plane starting from the sphere \cite{Dohi:2010vc}. For example, in \cite{Cacciapaglia:2009pa}, the parity assuring the stability of the dark matter candidate is due to a remnant of the 6-dimensional Lorentz symmetry after compactification, as the model has no fixed points (see \cite{Arbey:2012ke} for a detailed discussion). There are many reasons to study UED models with 2 compact extra dimensions; primarily as they may provide a dark matter candidate, suppress the proton decay rate, as well as providing anomaly cancellations for the number of light chiral fermion generations being a multiple of three \cite{Dobrescu:2001ae}.

\par Another important point connected to BSM physics, is that in the SM the neutrino does not have a mass, but neutrino oscillations suggest that neutrinos have finite mass and lepton flavours mix. The most recent evidence is the measurement of a large $\theta_{13}$ mixing parameter \cite{Abe:2011fz,An:2012eh,Ahn:2012nd}. In the context of UED models \cite{Bhattacharyya:2002nc}, one can use the dimension-five Weinberg operator  \cite{Weinberg:1979sa} to give Majorana masses to neutrinos and study RGEs for the physical observables in this sector. In general two extra dimensional models have different (and faster) evolution properties with respect to one extra dimensional models. It is therefore interesting to check if signs of the evolution of neutrino parameters are within experimental reach or not. 

\par In section \ref{sec:quartic} we introduce the Higgs quartic coupling RGEs and compare the limits on the effective theory obtained by requiring the stability of the Higgs potential with other effective rules for the cutoff theory, obtained from other requirements such as perturbativity of the interactions, gauge coupling unification, unitarity. In section \ref{sec:neutrino} we shall study the neutrino mixing and masses evolution and compare them with the quark sector RGEs discussed in \cite{Abdalgabar:2013oja}. Our conclusions will be given in section \ref{sec:concl}.


\section{The Quartic coupling RGEs}\label{sec:quartic}

\par We study a generic model with two universal extra dimensions, where in the following we summarise the evolution equations. We shall use a notation similar to the ones of \cite{Abdalgabar:2013oja,Cornell:2010sz}, noting that the beta functions contain terms quadratic in the cut-off, where this part dominates the evolution of the Yukawa couplings and of $k$ (the coefficient of the Weinberg operator). The top Yukawa coupling becomes non-perturbative only after the gauge coupling unification. As such, the limit on the range of validity for the effective theory will be determined by which condition is reached first: unification of the gauge couplings or instability of the Higgs potential. We shall first write down the results of the SM and then generalise it to include the effects arising from the extra dimensional degrees of freedom. The initial values we shall adopt at the $M_Z$ scale are: for the gauge couplings $\alpha_1(M_Z) = 0.01696$, $\alpha_2(M_Z) = 0.03377$, and $\alpha_3(M_Z) = 0.1184$, and for the fermion masses $m_u(M_Z) = 1.27$ MeV, $m_c(M_Z) = 0.619$ GeV, $m_t (M_Z) = 173.3$ GeV, $m_d(M_Z) = 2.90$ MeV, $m_s(M_Z) = 55$ MeV, $m_b (M_Z) = 2.89$ GeV, $m_e(M_Z) = 0.48657$ MeV, $m_\mu(M_Z) = 102.718$ MeV, and $m_\tau(M_Z) = 1746.24$ MeV \cite{Xing:2007fb,Cornell:2011fw}.

\subsection{SM evolution equations}

\par The evolution equations for the SM are a limiting case when the Kaluza-Klein (KK) scale becomes heavy and the KK modes decouple. We introduce them to fix the notation and as they are relevant below the KK threshold. When $0 < t < \ln (\frac{1}{{{M_Z}R}})$, where $t = \mathrm{ln}\left( \frac{\mu}{M_Z}\right)$, $\mu$ being the energy, that is, for the evolution between ${M_Z} < \mu < 1/R$; the Yukawa evolution equations are dictated by the usual SM ones:
 \begin{eqnarray}
\beta_{u}^{SM} &=& Y_u \left[-8g_3^2 -\frac{9}{4}g^2_2 -\frac{17}{20} g^2_1+\frac{3}{2} (Y^{\dagger}_u Y_u-Y^{\dagger}_d Y_d) + Tr (3Y^{\dagger}_u Y_u+3Y^{\dagger}_d Y_d+Y^{\dagger}_e Y_e) \right]\; ,\\
\beta_{d}^{SM} &=&  Y_d \left[-8g_3^2 -\frac{9}{4}g^2_2 -\frac{1}{4} g^2_1+\frac{3}{2} (Y^{\dagger}_d Y_d-Y^{\dagger}_u Y_u)  + Tr (3Y^{\dagger}_u Y_u+3Y^{\dagger}_d Y_d+Y^{\dagger}_e Y_e) \right] \; , \\
\beta_{e}^{SM} &=&  Y_e \left[ -\frac{9}{4}g^2_2 -\frac{9}{4} g^2_1+\frac{3}{2} Y^{\dagger}_e Y_e  + Tr (3Y^{\dagger}_u Y_u + 3Y^{\dagger}_d Y_d+Y^{\dagger}_e Y_e) \right] \;, \\
\beta_{\lambda}^{SM} &=&  \left[ 12 \lambda^2-\left(\frac{9}{5}g^2_1 +9 g^2_2\right)\lambda+\frac{9}{4}\left(\frac{3}{25}g_1^4 +g_2^4 + \frac{2}{5} g^2_1 g^2_2\right)\right. \nonumber\\
&&\hspace{2cm}+\left. 4 \lambda Tr (3Y^{\dagger}_u Y_u+3Y^{\dagger}_d Y_d+Y^{\dagger}_e Y_e) - 4 Tr \left(3(Y^{\dagger}_u Y_u)^2+3(Y^{\dagger}_d Y_d)^2+(Y^{\dagger}_e Y_e)^2\right) \right]\; , \\
\beta_{k}^{SM} &=&  \left[ \left(-3g^2_2 +\lambda + 2 Tr (3Y^{\dagger}_u Y_u+3Y^{\dagger}_d Y_d+Y^{\dagger}_e Y_e) \right)k-\frac{3}{2} \left(k Y^{\dagger}_e Y_e + (Y^{\dagger}_e Y_e)^T k\right) \right] \; . \label{Beta-Yukawa-lambda-k_SM}
\end{eqnarray}
The notations are as follows: $g_1$, $g_2$, $g_3$ are respectively the $U(1)$, $SU(2)$, $SU(3)$ gauge couplings; $Y_i$ are the Yukawa coupling matrices where the index refers to the corresponding sector ($u$ for up-type, $d$ for down-type and $e$ for leptons); $\lambda$ is the Higgs quartic coupling and $k$ the coefficient of the Weinberg operator. These equations are modified when we enter the energy regime where the effects of the extra dimensions set in. The modifications depend on the particles non-decoupled at that energy scale and on the structure of the model. We shall consider two cases, one in which all particles can propagate in the extra dimensions (bulk case) and the other in which SM particles are constrained to the brane (brane case). Note that for completeness the runnings of the gauge coupling constants, as we have previously presented for these models \cite{Abdalgabar:2013oja}, have been given in Appendix \ref{app:0}.

\subsection{The 2UED scenarios}

\par The RGEs for the Yukawa couplings, Higgs quartic couplings and neutrino running parameter in the 2UED model, for all three generations propagating in the bulk, can be expressed as:
 \begin{eqnarray}
\beta_{u}^{6D} &=&  \pi(S(t)^2-1)Y_u \left[-\frac{32}{3}g_3^2 -\frac{3}{2}g^2_2 -\frac{5}{6} g^2_1+3 (Y^{\dagger}_u Y_u-Y^{\dagger}_d Y_d)+ 2 Tr (3Y^{\dagger}_u Y_u+3Y^{\dagger}_d Y_d+Y^{\dagger}_e Y_e) \right]\; , \\
\beta_{d}^{6D} &=& \pi(S(t)^2-1) Y_d \left[-\frac{32}{3}g_3^2 -\frac{3}{2}g^2_2 -\frac{1}{30} g^2_1+3 (Y^{\dagger}_d Y_d-Y^{\dagger}_u Y_u) + 2 Tr (3Y^{\dagger}_u Y_u+3Y^{\dagger}_d Y_d+Y^{\dagger}_e Y_e) \right] \; , \\
\beta_{e}^{6D} &=& \pi(S(t)^2-1) Y_e \left[ -\frac{3}{2}g^2_2 -\frac{27}{10} g^2_1+3 Y^{\dagger}_e Y_e + 2 Tr (3Y^{\dagger}_u Y_u+3Y^{\dagger}_d Y_d+Y^{\dagger}_e Y_e) \right] \; , \\
\beta_{\lambda}^{6D} &=& \pi(S(t)^2-1)  \left[ 12 \lambda^2-\left(\frac{9}{5}g^2_1 +9 g^2_2\right)\lambda+\left(\frac{9}{20}g_1^4 +\frac{15}{4}g_2^4 +\frac{3}{2} g^2_1 g^2_2\right)\right. \nonumber\\
&& \hspace{2.5cm}+\left. 8 \lambda Tr (3Y^{\dagger}_u Y_u+3Y^{\dagger}_d Y_d+Y^{\dagger}_e Y_e) - 8 Tr \left(3(Y^{\dagger}_u Y_u)^2+3(Y^{\dagger}_d Y_d)^2+(Y^{\dagger}_e Y_e)^2\right) \right]\; , \label{LambdaBulk} \\
\beta_{k}^{6D} &=& \pi(S(t)^2-1) \left[ \left(-\frac{3}{20}g^2_1 -\frac{5}{2} g_2^2+\lambda + 4 Tr (3Y^{\dagger}_u Y_u+3Y^{\dagger}_d Y_d+Y^{\dagger}_e Y_e) \right) k-3 (k Y^{\dagger}_e Y_e + (Y^{\dagger}_e Y_e)^T k) \right] \;.
\label{Beta-Yukawa-Bulk}
\end{eqnarray}

\par The corresponding evolution equations, for all three generations restricted to the brane are given by:
 \begin{eqnarray}
\beta_{u}^{6D} &=& 4\pi(S(t)^2-1)Y_u \left[-8g_3^2 -\frac{9}{4}g^2_2 -\frac{17}{20} g^2_1+\frac{3}{2} (Y^{\dagger}_u Y_u-Y^{\dagger}_d Y_d) \right]\; ,\\
\beta_{d}^{6D} &=&  4\pi(S(t)^2-1) Y_d \left[-8g_3^2 -\frac{9}{4}g^2_2 -\frac{1}{4} g^2_1+\frac{3}{2} (Y^{\dagger}_d Y_d-Y^{\dagger}_u Y_u)  \right] \; , \\
\beta_{e}^{6D} &=& 4\pi(S(t)^2-1) Y_e \left[ -\frac{9}{4}g^2_2 -\frac{9}{4} g^2_1+\frac{3}{2} Y^{\dagger}_e Y_e  \right] \;,\\
\beta_{\lambda}^{6D} &=& \pi(S(t)^2-1) \left[ 12 \lambda^2-\left(\frac{9}{5}g^2_1 +9 g^2_2\right)\lambda+\left(\frac{9}{20}g_1^4 +\frac{15}{4}g_2^4 +\frac{3}{2} g^2_1 g^2_2\right) \right]\; ,\label{LambdaBrane} \\
\beta_{k}^{6D} &=& 2\pi(S(t)^2-1) \left[ (-3 g_2^2+\lambda )\, k -3 \left(k Y^{\dagger}_e Y_e + (Y^{\dagger}_e Y_e)^T k\right) \right] \;, \label{Beta-Yukawa-Brane}
\end{eqnarray}
where $S(t)= M_Z R e^t$, assuming that all modes contribute in the range of our energy scale for $t = \mathrm{ln}(\mu/M_Z)$. Note that these coefficients are model dependent, as discussed further in Appendix \ref{app:A}. The bulk and brane sets of evolution equations share the same structure but bring about quite different evolutions for the physical parameters, for example if you compare Eq. (\ref{LambdaBulk}) for the bulk case against the corresponding one for the brane case Eq. (\ref{LambdaBrane}) you can see that the presence of the Yukawa terms adds a negative contribution which will affect the evolution. Numerically we will show in the following that this term is dominant and drives the quartic coupling to zero in the bulk case as the energy scale increases, whilst in the brane model, which does not contain such a contribution, has the opposite behaviour and the quartic coupling grows with the energy scale. Our calculation agrees with \cite{Ohlsson:2012hi} in the general structure, but the  coefficient of $g^2_1$ in the running of the $k$ parameter is different and the number of the Kaluza-Klein  particles taken into account in the factor $S^2(t)$ is also different (see Appendix \ref{app:A}). In particular we have explicitly calculated the KK modes contributing up to the cut-off in different 6D models, while \cite{Ohlsson:2012hi} only has a factor of two with respect to the 5D case. Using a factor of two amounts to considering only the modes $(j,0)$ and $(0,k)$, while disregarding all the ``mixed" modes $(j,k)$ with $j,k\neq 0$. Even if the numerical differences are not very large, excluding the mixed modes is inconsistent.

\subsection{The 2UED bulk and brane quartic results}

\begin{figure}[h]
\begin{center}
\includegraphics[width=7cm,angle=0]{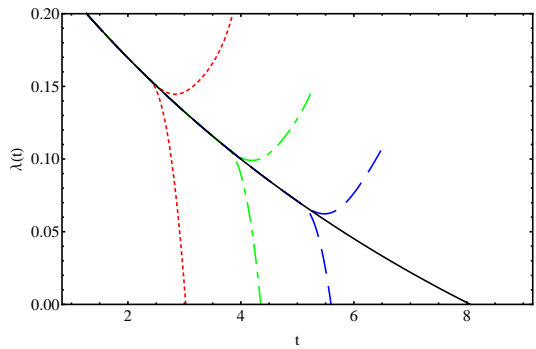} \qquad
\end{center}
\caption{\it (Colour online) The evolution of the Higgs quartic coupling {$\lambda$}, where the solid line represents the SM case with: downward trajectories for all matter fields in the bulk; and upward for all matter fields on the brane; for three different values of the compactification scales 1 TeV (dotted line), 4 TeV (dot-dashed line) and 10 TeV (dashed line), as a function of the scale parameter {$t$}.}
\label{lambda-2ued}
\end{figure}
\begin{figure}[h]
\begin{center}
\includegraphics[width=7cm,angle=0]{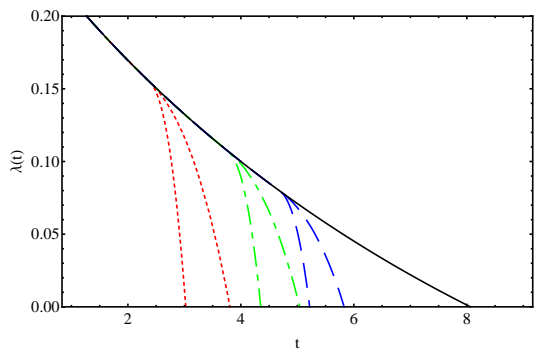} \qquad
\end{center}
\caption{\it (Colour online) Comparison of the Higgs quartic coupling evolution between the 1UED case and the 2UED case, where the solid line represent the SM case; all matter fields are in the bulk; for a compactification scale of 1 TeV  (dotted line), 4 TeV (dot-dashed line) and 10 TeV (dashed line), as a function of the scale parameter {$t$}. The 2UED line is always steeper than the corresponding 1UED one.}
\label{lambda-comparison}
\end{figure}

\par The numerical calculation of the RGEs confirms the results expected by the inspection of the analytical formulae above for the brane and for the bulk UED models. We show in Fig. \ref{lambda-2ued} the evolution of the Higgs quartic coupling in these two scenarios (all matter fields propagate in the bulk (downward evolution with increasing energy scale) or brane localised matter fields (upward evolution with increasing energy scale)). As can be seen, for the brane case the quartic Higgs coupling $\lambda (t)$ is positive and remains finite as we run from the EW scale all the way up to the unification scale. In contrast to the bulk case (and also in the SM) the evolution of $\lambda (t)$ goes to zero at some energy scale before reaching the unification scale, which implies the vacuum instability of the model and requires the introduction of a cut-off which is typically lower than the one usually determined by other means. A discussion of this point was already performed in the literature for the bulk case in \cite{Ohlsson:2012hi} for a particular case (assuming that the number of modes in the 6D case is twice the one of the corresponding 5D model). Our numerical results agree qualitatively with theirs, as the running has a similar behaviour, but we consider realistic models taking into account explicitly all the KK modes up to the cut-off. Recall, as defined earlier in section \ref{sec:quartic}, our effective theories cut-off is determined by either the reaching of the gauge coupling unification (see Appendix \ref{app:0}) or instability of the Higgs potential.

\par More general results were also obtained in \cite{Kakuda:2013kba} for 2UED results. We use the updated experimental values for the Higgs sector from the ATLAS and CMS collaborations and updated values for the top quark mass and also consider more general bulk and brane scenarios. We have checked that the dependence on the Higgs boson and top quark masses in their experimentally allowed ranges does not affect significantly the result of the evolution. The evolution is also only weakly sensitive to the particular choice of 2UED model (we have considered the two broad classes of models issued from the compactifications of the crystallographic groups of the plane and of the sphere $S^2$). A more detailed discussion of the model dependence of the results is given in Appendix \ref{app:A}.

\par In Fig.\ref{lambda-comparison} we present a comparison evolution of Higgs quartic couplings in the bulk case between the 1UED and 2UED model for different values of compactification scale (1, 4 and 10 TeV). We find that the evolution has the same behaviour, but in the 2UED model the cut-off is lower than the 1UED model, this being due to the presence of $S^2 (t)$ in equation (\ref{Beta-Yukawa-Bulk}) instead of the linear dependence on $S(t)$ as in the 1UED model. 


\section{Neutrino mixing and masses}\label{sec:neutrino}

\par We first state our conventions for the mixing angles and phases and briefly discuss different scenarios for neutrino masses. The mixing matrix which relates gauge and mass eigenstates is defined to diagonalise the neutrino mass matrix in the basis where the charged lepton mass matrix is diagonal \cite{Maki:1962mu}:
\begin{equation}
U = \left( \begin{array}{ccc}
c_{12} c_{13} & s_{12} c_{13} & s_{13} e^{- i \delta} \\
-s_{12} c_{23} - c_{12} s_{23} s_{13} e^{- i \delta} & c_{12} c_{23} - s_{12} s_{23} s_{13} e^{i \delta} & s_{23} c_{13} \\
s_{12} s_{23} - c_{12} c_{23} s_{13} e^{i \delta} & - c_{12} s_{23} - s_{12} c_{23} s_{13} e^{i \delta} & c_{23} c_{13}
\end{array} \right)
\left( \begin{array}{ccc}
e^{i \phi_1} && \\
& e^{i \phi_2} & \\
&& 1
\end{array} \right) \; , \nonumber
\end{equation}
with $c_{ij} = \cos \theta_{ij}$ and $s_{ij} = \sin \theta_{ij}$ ($ij = 12, 13, 23$).  We follow the conventions of \cite{Antusch:2003kp} 
to extract mixing parameters from the PMNS matrix.

\begin{center}
\begin{table}[h!]
\begin{tabular}{c|c}
Parameter & Value (90\% CL) \\ \hline
$\sin^2(2\theta_{12})$ & $0.861(^{+0.026}_{-0.022})$ \\
$\sin^2(2\theta_{23})$ & $>0.92$ \\
$\sin^2(2\theta_{13})$ & $0.092\pm0.017 $ \\
$\Delta m^2_{\mathit{sol}}$ & $(7.59\pm 0.21)\times 10^{-5}eV^2$ \\
$\Delta m^2_{\mathit{atm}}$ & $(2.43\pm 0.13)\times 10^{-3}$ $eV^2$ \\
\end{tabular}
\caption{\it Present limits on neutrino masses and mixing parameters. Data is taken from \protect{\cite{An:2012eh}} for sin$^2(2\theta_{13})$, and from \protect{\cite{Beringer:1900zz}}.}
\label{tableexp}
\end{table}
\end{center}

\par Experimental information on neutrino mixing parameters and masses is obtained mainly from oscillation experiments \cite{Nakamura:2010zzi,Beringer:1900zz}. In general $\Delta m^2_{\mathit{atm}}$ is assigned to a mass difference between $\nu_3$ and $\nu_2$, whereas $\Delta m^2_{\mathit{sol}}$ to a mass difference between $\nu_2$ and $\nu_1$. The current observational values are summarised in Table~\ref{tableexp}. Data indicates that $\Delta m^2_{\mathit{sol}} \ll \Delta m^2_{\mathit{atm}}$, but the masses themselves are not determined. In this work we have adopted the masses of the neutrinos at the $M_Z$ scale as $m_{1} = 0.1$ eV, $m_{2} = 0.100379$ eV, and $m_{3} = 0.11183$ eV, as the {\it normal} hierarchy (whilst any reference to an {\it inverted} hierarchy would refer to $m_{3} = 0.1$ eV, with $m_{3} < m_{1} < m_{2}$ and satisfying the above bounds). For the purpose of illustration, we choose values for the angles and phases at the $M_Z$ scale as: $\theta_{12} = 34^o$, $\theta_{13} = 8.83^o$, $\theta_{23} = 46^o$, $\delta =30^0$, $\phi_1 = 80^o$ and $\phi_2 = 70^o$.

\par Using the 2UED model \cite{Cornell:2012uw}, the transition to the 2UED bulk case will be done by making the replacement of 
$C=\pi(S(t)^2-1)$ and $\alpha=\pi(S(t)^2-1) \left[ -\frac{9}{10}g^2_1 -\frac{5}{2} g_2^2+\lambda + 4 Tr (3Y^{\dagger}_u Y_u+3Y^{\dagger}_d Y_d+Y^{\dagger}_e Y_e)  \right]$ in Eqs. (B1--D3) in \cite{Cornell:2012uw}, and Eq. (A.3) in \cite{Cornell:2012qf}. Similarly, we will also have the same equations in the 2UED brane case, with $C=2\pi (S(t)^2-1)$ and $\alpha=2\pi (S(t)^2-1) (-3 g_2^2 + \lambda)S(t)$.

\par
\begin{figure}[h]
\begin{center}
\includegraphics[width=7cm,angle=0]{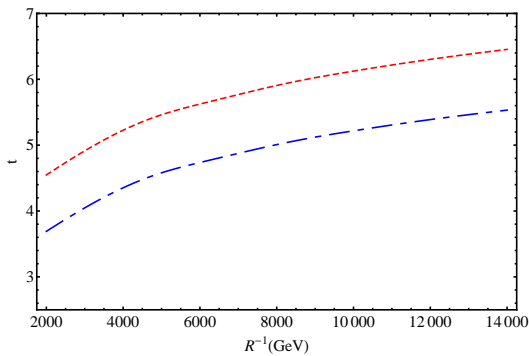} \qquad
\end{center}
\caption{\it (Colour online) The cut-off of the Higgs quartic coupling {$\lambda$} (dot-dashed line) and gauge couplings (dashed line) for all matter fields in the bulk; for different values of the compactification scales from 2 TeV to 14 TeV, as a function of the energy scale parameter {$t$}.}
\label{lambda_gauge_cutoff-2ued}
\end{figure}

 In Fig. \ref{lambda_gauge_cutoff-2ued} we plot the cut-off of the Higgs quartic coupling and gauge couplings for all matter fields propagating in the bulk. As we observed before in Fig. \ref{lambda-2ued} for the bulk case, the Higgs self-couplings evolves towards zero at high energies requiring the introduction of an ultraviolet cut-off for the theory. As can be seen from the plot, the cut-off required by the $\lambda$ evolution reaching zero is lower than the gauge couplings unification scale. 

\par
\begin{figure}[h]
\begin{center}
\includegraphics[width=7cm,angle=0]{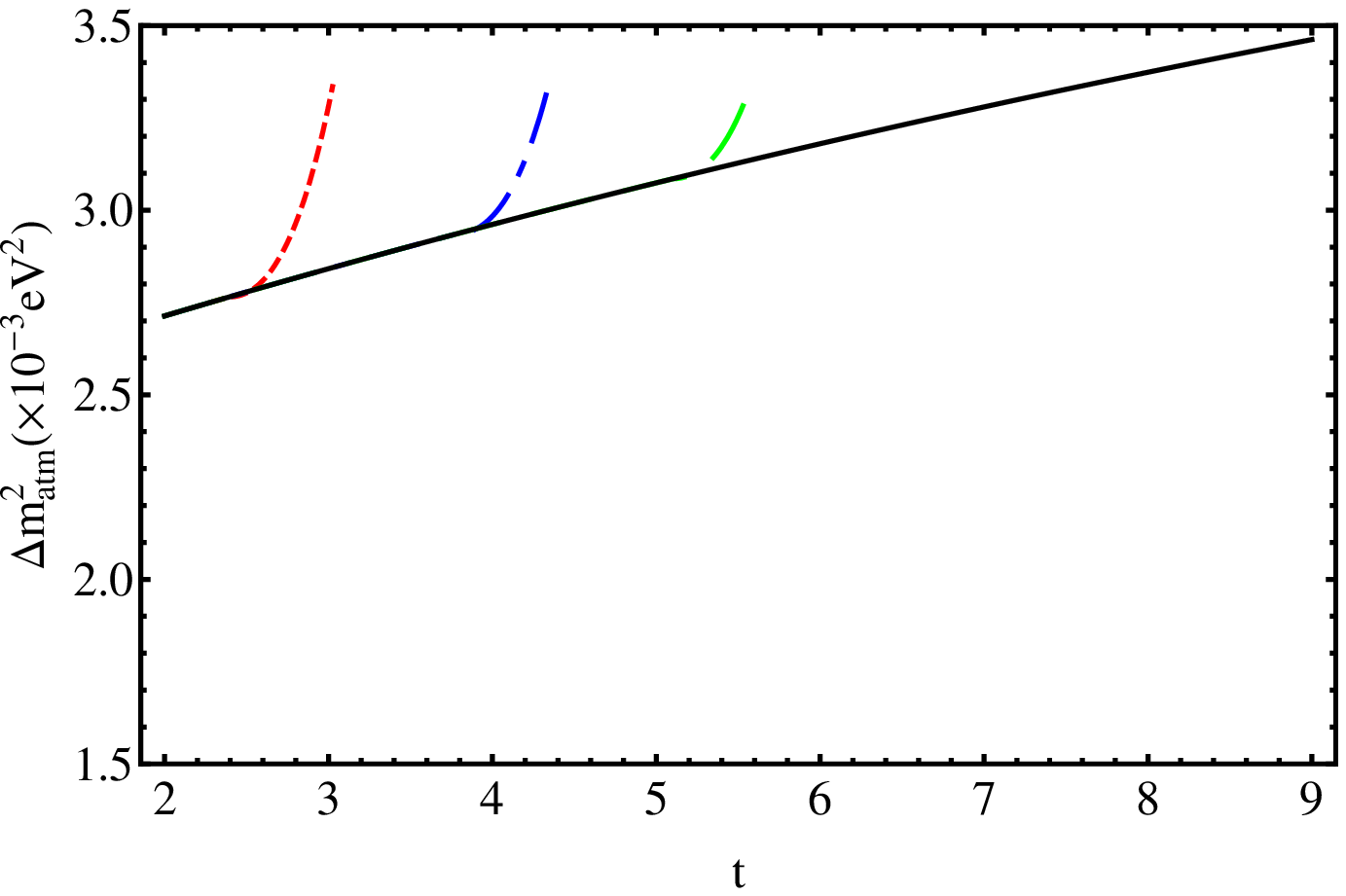} \qquad
\includegraphics[width=7cm,angle=0]{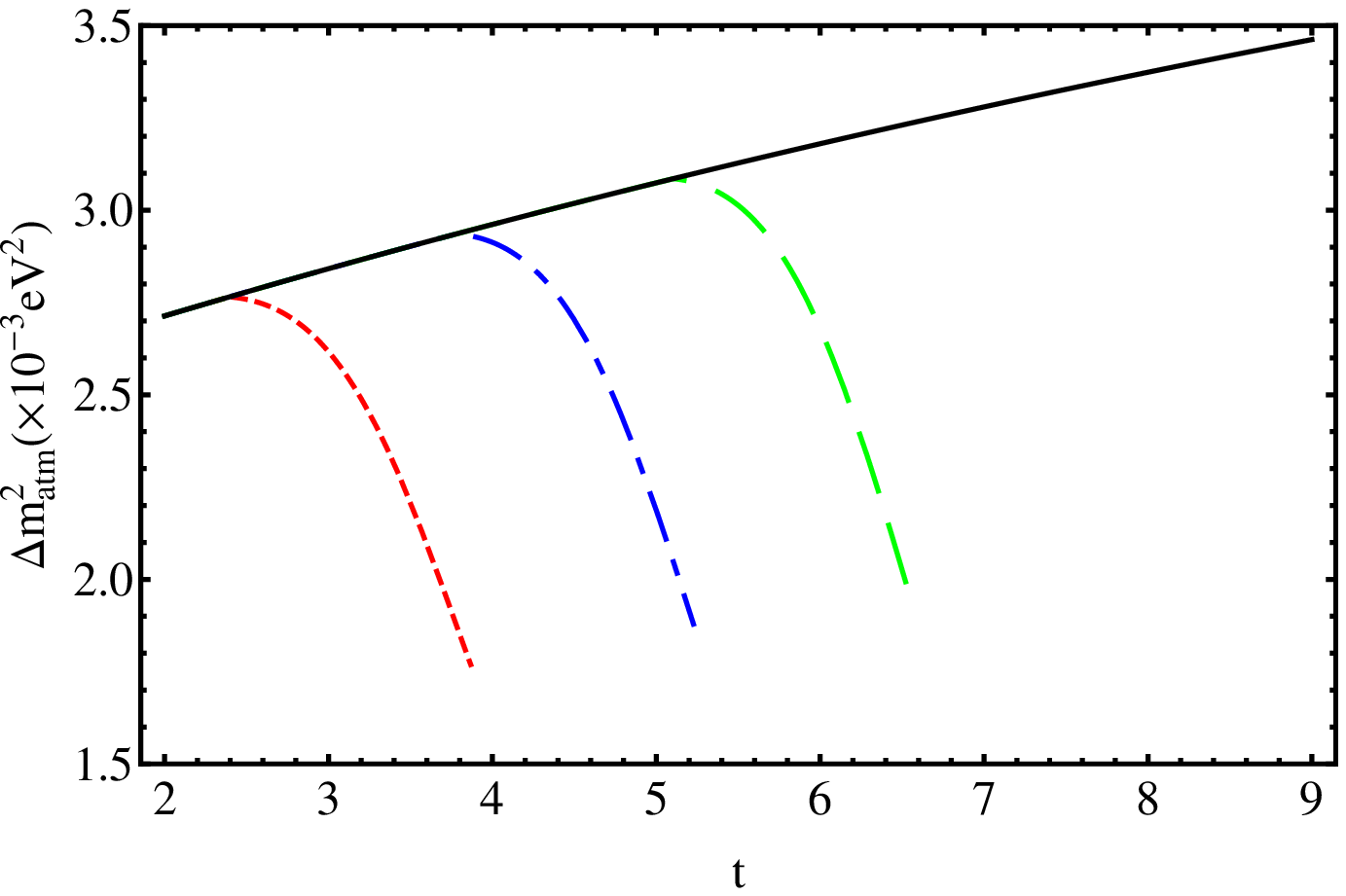}
\end{center}
\caption{\it (Colour online) The evolution of $\Delta m^2_{atm}$ where the solid line represents the SM case with: in the left panel all 
matter fields in the bulk; and the right panel for all matter fields on the brane; for three different values of the compactification scales 1 
TeV (dotted line), 4 TeV (dot-dashed line) and 10 TeV (dashed line), as a function of the scale parameter {$t$}.}
\label{dmatm-2ued}
\end{figure}

 The evolution of the mass squared differences $\Delta m^2_{atm}$  both for the matter fields on the brane and for all fields in the bulk is presented in Fig. \ref{dmatm-2ued}. Only some selected plots will be shown and we will comment on the other similar cases not explicitly shown. As depicted in Fig. \ref{dmatm-2ued} the mass squared difference increases rapidly once the KK threshold is crossed at $\mu= R^{-1}$ for the bulk case, resulting in a rapid approach to a singularity before the unification scale is reached, note however that  the cut-off imposed by the requirement of stability of the Higgs potential is reached much faster. For the brane localised case the contribution from the gauge couplings is important, and the evolution decrease instead of increasing. Note that $\Delta m^2_{sol}$ has the same shape as  $\Delta m^2_{atm}$ for both cases. To see the running behaviours of neutrino mixing parameters in the 2UED model, we carry out similar numerical analyses by using the beta function derived in section \ref{sec:quartic} for both the bulk case and the brane case. From this we observe that the correction to $\theta_{13}$ and $\theta_{23}$ are quite small and milder than $\theta_{12}$. For $\theta_{12}$ the largest variations are of the order of $0.3\%$ on the full energy range of validity for the effective theory. For $\theta_{13}$ and $\theta_{23}$ the variations are negligible. There is therefore no substantial difference in all cases with respect to the SM.

\section{Conclusions}\label{sec:concl}

\par In this work we derived the RGEs for the Higgs quartic coupling and neutrino mass running for two distinct classes of 2UED models, that of all matter fields propagating in the bulk or localised to the brane.  We obtain stronger constraints on the cut-off scale from the requirement of the stability of the Higgs potential in the bulk case. Whilst in the brane case the evolution of the quartic Higgs coupling has improved vacuum stability and $\lambda$ is positive and finite from the EW scale all the way up to the unification scale. We also compare our results with the 1UED model, where we find a more rapid evolution of the physical observables in the 2UED models. 

\par On the other hand, in the neutrino sector, the evolution equations for the mixing angles, phases, and  $\Delta m^2_{atm}$ and $\Delta m^2_{sol}$ are also considered. Once the first KK threshold is reached, these quantities increase with increasing energy for the bulk case and decrease with energy in the brane case.  However the effect is almost negligible for the mixing angles, while it can be sizeable for the evolution of the squared mass differences.


\section*{Acknowledgements}

This work is supported by the National research Foundation (South Africa) and by Campus France (Project Protea-29719RB). We also acknowledge partial support from the Labex-LIO (Lyon Institute of Origins).

\appendix 


\section{Gauge couplings evolution}\label{app:0}

\par The gauge coupling RGE for all matter fields propagating in the bulk \cite{Abdalgabar:2013oja}, where apart from the SM field contributions, there will be new contributions from the spinless adjoints $A^{(j,k)}_H$. As such, the calculation is similar to that of the 5D UED model but with an additional  factor of 2 due to 6D gauge field having two extra dimensional components. Note that for the case of all matter fields being restricted to the brane there will be no contributions from the KK excited states of the fermions. The generic structure of the one-loop RGE for the gauge couplings is then given by:
\begin{equation}
16 \pi^2 \frac{d g_i}{d t}= b^{SM}_i g^3_i+\pi \left( S(t)^2-1 \right) b^{6D}_i g^3_i \;,
\label{gauge2UED}
\end{equation}
where $t = \ln (\frac{\mu}{M_Z})$, $S(t) = {e^t}{M_Z}R$, or $S(\mu)=\mu R=\frac{\mu}{M_{KK}}$ for $M_Z < \mu < \Lambda$ ($\Lambda$ is the cut-off scale. The numerical coefficients appearing in equation (\ref{gauge2UED}) are given by:
\begin{equation}
b^{SM}_i=  \left[ \frac{41}{10}, -\frac{19}{6}, -7\right]\;,
\end{equation}
and 
\begin{equation}
b^{6D}_i= \left[\frac{1}{10}, -\frac{13}{2}, -10\right]+\left[\frac{8}{3}, \frac{8}{3}, \frac{8}{3}\right]\eta\;,
\end{equation}
\noindent $\eta$ being the number of generations of fermions propagating in the bulk. Therefore, in the two cases we shall consider, that of all fields propagating in the bulk ($\eta =3$) we have \cite{Cornell:2011fw}:
\begin{equation}
b^{6D}_i= \left[\frac{81}{10}, \frac{3}{2}, -2\right]\;.
\end{equation}
\noindent Similarly, for all matter fields localised to the brane ($\eta =0$) we have:
\begin{equation}
b^{6D}_i= \left[\frac{1}{10}, -\frac{13}{2}, -10\right]\;.
\end{equation}

\begin{figure}[h]
\begin{center}
\includegraphics[width=7cm,angle=0]{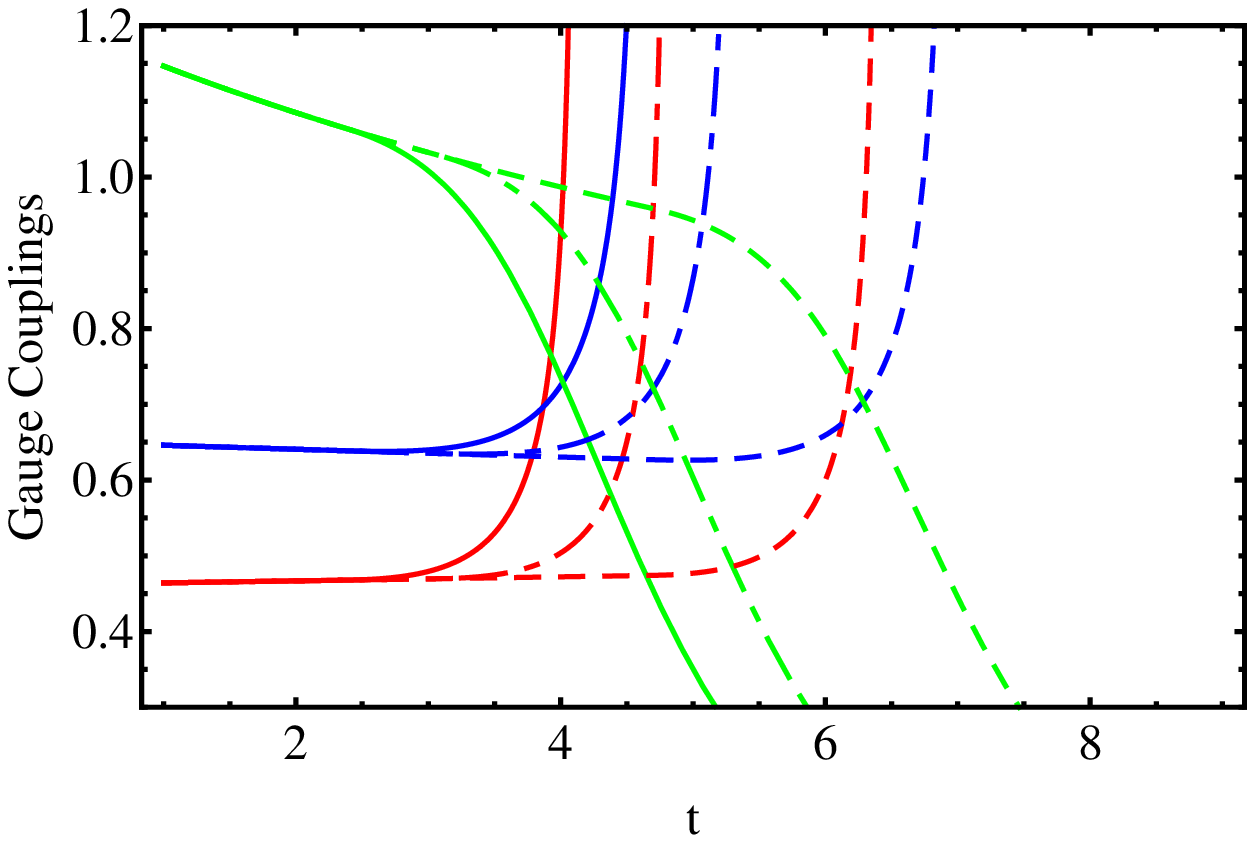} \qquad
\includegraphics[width=7cm,angle=0]{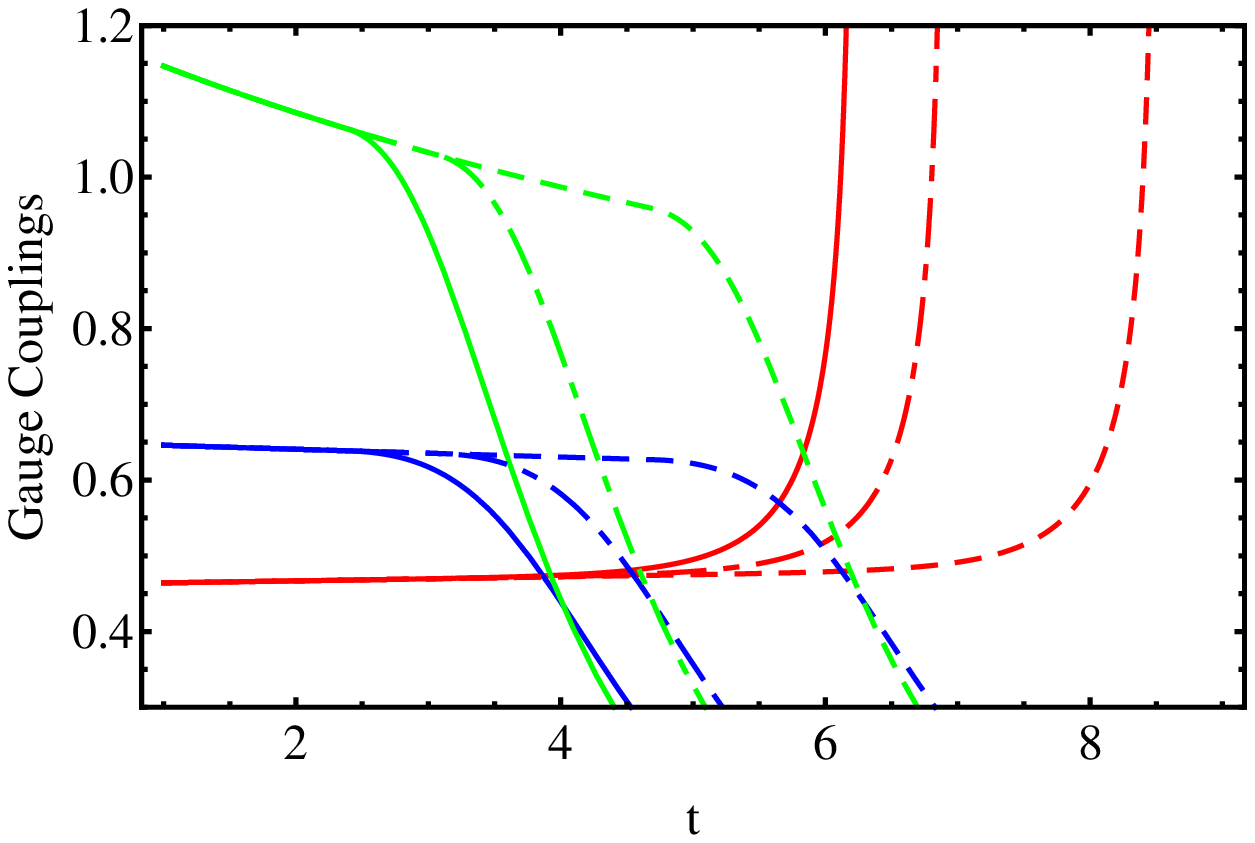}
\end{center}
\caption{ \it (Colour online) The evolution of gauge couplings {$g_1$} (red), {$g_2$} (blue) and {$g_3$} (green), with: in the left panel, all matter fields in the bulk; and the right panel for all matter fields on the brane; for three different values of the compactification scales 1 TeV (solid line), 2 TeV (dot-dashed line) and 10 TeV (dashed line), as a function of the scale parameter {$t$}.}
\label{gauge-2ued}
\end{figure}
\begin{figure}[h]
\begin{center}
\includegraphics[width=7cm,angle=0]{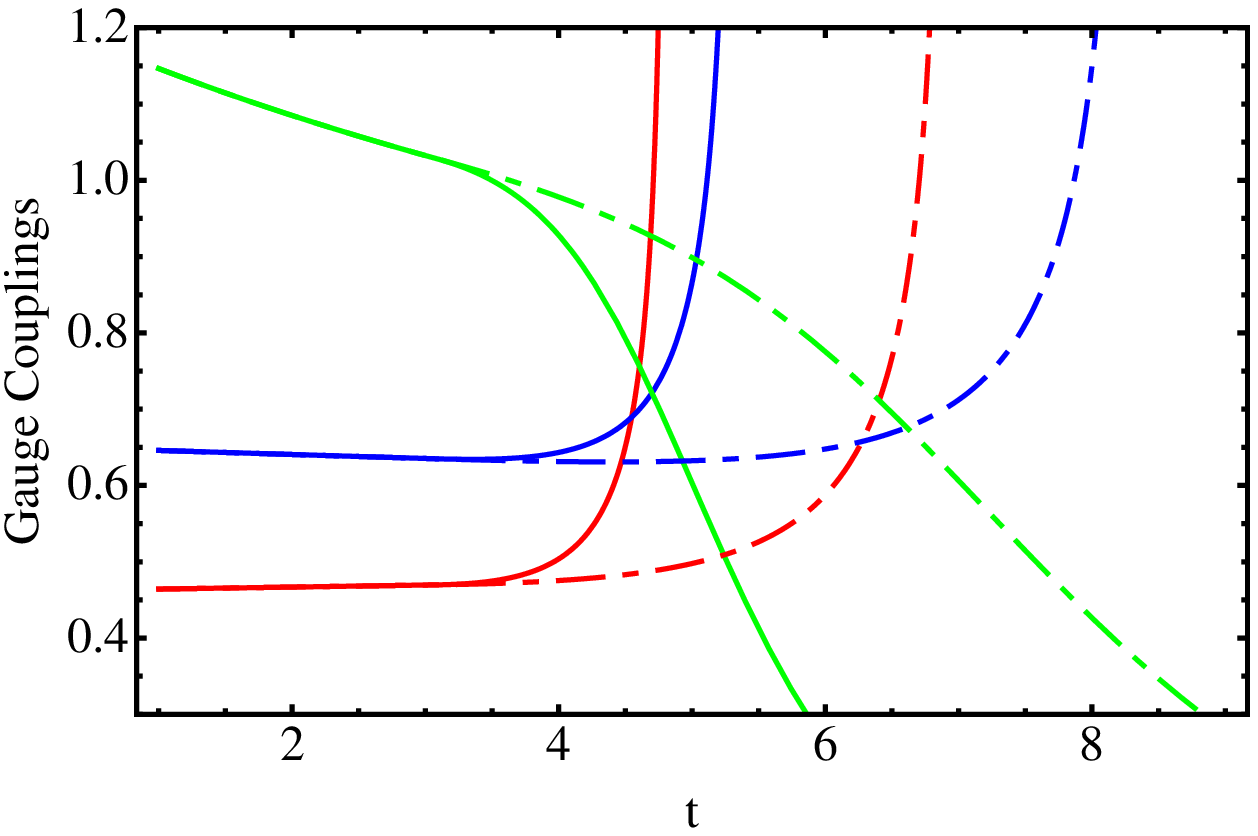} \qquad
\includegraphics[width=7cm,angle=0]{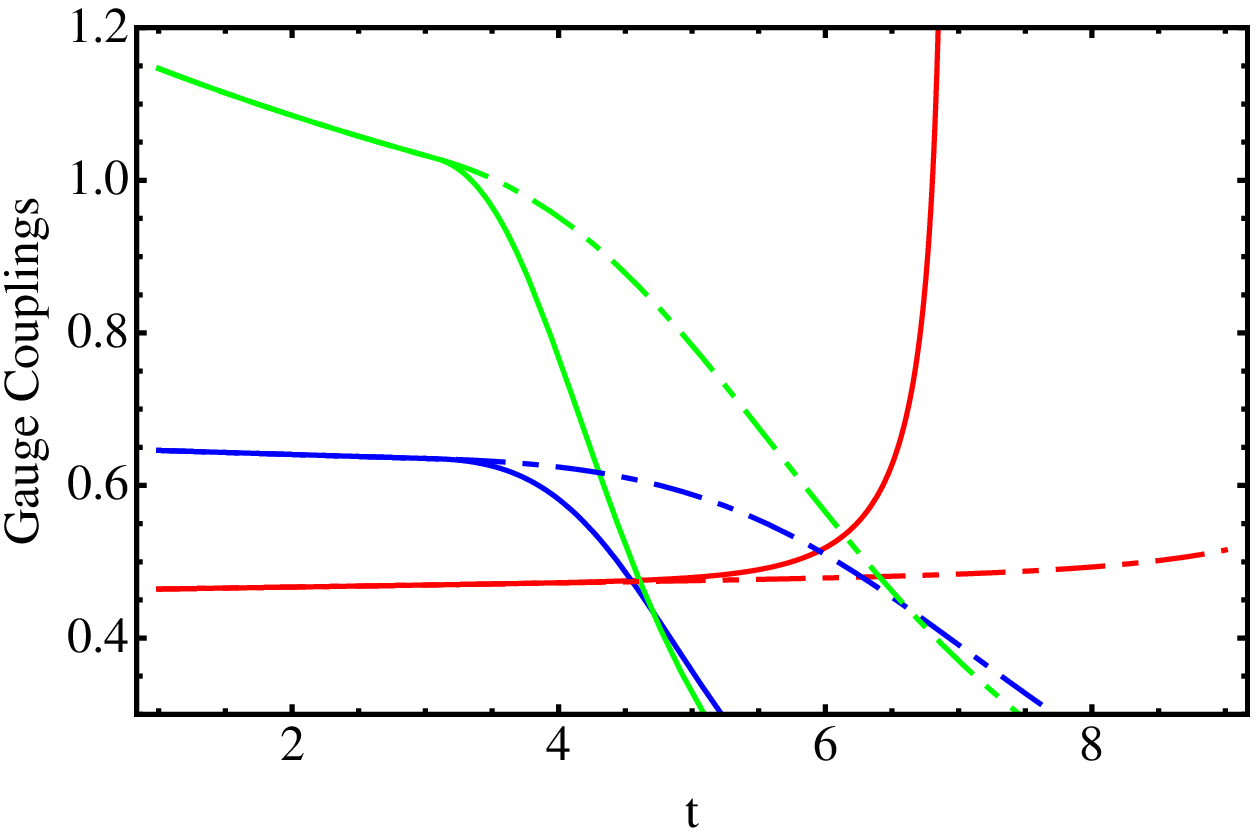}
\end{center}
\caption{ \it (Colour online) Comparison of the gauge coupling evolutions {$g_1$} (red), {$g_2$} (blue), {$g_3$} (green) between the 1UED case (dashed line) and the 2UED case (solid line) with: in the left panel, all matter fields in the bulk; and the right panel for all matter fields on the brane; for a compactification scale of 2 TeV as a function of the scale parameter {$t$}.}
\label{gauge-comparison}
\end{figure}

\par We present in Fig.\ref{gauge-2ued} the evolution of the bulk field and brane localised cases for several choices of compactification scale for the extra-dimension in the 2UED model. We find that there is a difference in the $g_2$ evolution, where it increases in the bulk propagating case and decreases in the brane localised case. We also see that the three gauge coupling constants, as expected in extra-dimensional theories, can unify at some value of $t$ depending on the radius of compactification. As an example, for 1 TeV we see an approximation unification at $t=4$. 
 
\par In Fig.\ref{gauge-comparison} we show for comparison the gauge couplings between the 1UED and 2UED cases for a compactification scale of 2 TeV. From the plots and the discussion in Ref.\cite{Cornell:2012qf}, we see that in both cases the gauge couplings have similar behaviour, however in the 2UED case we have asymptotes at lower $t$ values, that is, a lower energy scale. As such the range of validity for the 2UED is less than the 1UED case, this being due to the $S^2(t)$ factor present in equation (\ref{gauge2UED}), there only being a linear dependence on $S(t)$ for the 1UED case.

\par The solid line (which corresponds to the 2UED case) drops off faster than the dashed line (1UED case) when the gauge couplings decrease with energy scale. For the $g_1$ coupling, it increases faster than in the 2UED case (at $t \sim 6$) with a roughly constant evolution in the 1UED case. As such one can see in the brane case a large difference in the evolution of this coupling, a feature which can distinguish these two models.

\par As such the appropriate cut-off for the three radii considered in this paper, for the brane localized matter fields case, will be determined by the instability of the Higgs quartic condition. This will correspond to $t(R^{-1} = $ 1 TeV$) \sim 3.0$, $t(R^{-1} = $ 4 TeV$) \sim 4.3$ and $t(R^{-1} = $ 10 TeV$) \sim 5.7$ (see Fig.\ref{lambda-comparison}). For the bulk case, the cut-off has been presented in table 1 of \cite{Abdalgabar:2013oja}. Note that this corresponds to approximately 5 KK modes in the 2UED $R^{-1} =$ 1 TeV case being accommodated before the cut-off is reached; these being the $(j, k)$ modes (1,0), (1,1), (2,0), (1,2) and (2,1) (note that (0,1) and (0,2) are excluded by the selection rules given in section 2 of \cite{Abdalgabar:2013oja}).


\section{Model dependence of the RGEs in 2UED models}\label{app:A}

\par Different 2UED models do not share exactly the same KK spectrum. A detailed calculation can be found in the Appendix of \cite{Abdalgabar:2013oja} to which we refer the interested reader. The KK number for the general 2UED model is $2C (S(t)^2 -1)$, where $C=\pi/2$, $S(t)= M_Z R e^t$ for our general reference model (the torus $T^2$), assuming that all modes contribute. Models based on the compactifications of the crystallographic groups of the plane are very similar to the case of the torus $T^2$. In the case of compactifications based on the sphere $S^2$, the KK coefficient, as function of the $t$ parameter, is given by $2(S(t)^2 -1)$. In specific realisations of the 2UED models some states are not present due to orbifold symmetry requirements on the wavefunctions, and one should subtract the states which do not contribute to the total KK coefficient. We have tested the model dependence of the results obtained with the RGEs and only minor changes were observed, while the conclusions and the phenomenology discussed in the text remain unaltered. 


\end{document}